\documentstyle[aas2pp4]{article}		

\received{}
\accepted{}
\journalid{504}{1 September 1998}

\lefthead{Church et al.}
\righthead{Spectral Evolution during Dips in XBT\thinspace 0748-676} 

\begin{document}

\centerline {\large {\it to be published in Astrophysical Journal vol.}
{\bf 504}{, }{\it September 1st, 1998}}

\vskip 0.5 cm

\title{Progressive Covering of the ADC during Dipping in the LMXB 
XBT\thinspace 0748-676}
\author{M. J. CHURCH\altaffilmark{1}, M. BA\L UCI\'NSKA-CHURCH\altaffilmark{1},
T. DOTANI\altaffilmark{2}, AND K. ASAI\altaffilmark{2}}

\altaffiltext{1}{School of Physics and Astronomy, University of
Birmingham, Edgbaston, Birmingham B15 2TT, UK}
\altaffiltext{2}{Institute of Space and Astronautical Science, Yoshinodai
3-1-1, Sagamihara, Kanagawa 229-8510, JAPAN}


\begin{abstract}
We report results of analysis of the ASCA observation of 1993, May 7th of the dipping
LMXB source XBT\thinspace 0748-676, and propose a new explanation of the spectral evolution 
in dipping in this source. The behavior of the source was very unusual in that, in 
the band 1 - 3 keV, dipping extended around most of the orbital cycle with almost no 
non-dip intensity evident and the depth of dipping reached 100\%. At higher energies, 
eg 3 - 10 keV, the depth of dipping was less than 100\%, and there were 
marked increases in hardness in dipping. 
We show that the non-dip and dip spectra in several 
intensity bands are well fitted using the same physical model that we have previously 
shown gives good explanations of several dipping sources, consisting of point source 
blackbody emission from the neutron star, plus extended Comptonised emission from the 
accretion disk corona (ADC), with progressive covering of the ADC during dipping.
Best fit values of $\rm {kT_{bb}}$ = 1.99$\pm $0.16 keV and power 
law photon index $\Gamma $ = 1.70$ \pm $0.16 are found. The strong excess
below 1 keV was well fitted by a Gaussian line at 0.65 keV. In dipping,
good fits were obtained by allowing it to be covered by the same progressive
covering factor as the extended continuum emission, providing strong
evidence that the line originates in the ADC.
Our approach of applying the 
two-component model, and explicitly including progressive covering of the Comptonised emission, 
differs radically from the ``absorbed plus unabsorbed'' approach previously used 
extensively for XBT\thinspace 0748-676 and similar sources, in which the normalisation 
of the unabsorbed peak in dip spectra is allowed to decrease by a large factor in dipping.
This decrease has often been attributed to the effects of electron scattering. By using 
our two-component model we show that the unabsorbed component is the uncovered fraction 
of the Comptonised emission, and in the band 1 - 10 
keV, we do not need to invoke electron scattering to explain dipping. 
\end{abstract}



\section{Introduction}
There are $\sim $ 10 Low Mass X-ray Binaries (LMXB) that exhibit absorption dips
in X-ray intensity at the orbital period, and it is generally accepted that 
these are due to absorption in the bulge in the outer accretion disk where
the accretion flow from the Companion impacts (White and Swank 1982). 
XBT\thinspace 0748-676 is an important member of the dipping class, discovered 
and observed extensively with {\t Exosat}, exhibiting dipping, bursting and
eclipsing (Parmar et al. 1986). Since then, it has undergone strong
variations in brightness. The {\it Exosat} observations showed that
both dips and interdips were present, and that the depth and extent of dipping 
varied markedly from orbit to orbit. Eclipses showed that the inclination was high,
and during these, about 4\% of the non-dip emission in the band 2 - 6 keV
remained, indicating the
presence of an extended emission component. Using the eclipses as fiducial markers,
Parmar et al. (1986, 1991) found a very accurate value for the orbital period
and a decreasing period. Asai et al. (1992) and
Corbet et al. (1994) showed evidence that the period is modulated
sunusoidally, but this pattern was not continued when RXTE data was added
(Hertz et al. 1997). However all determinations
agree on the period to $\sim $ 0.01~s, ie 13766.78~s.

\medskip \noindent
The dipping sources do not, in general, have the spectral evolution expected 
for photoelectric absorption of a single emission component in the bulge in the outer
accretion disk, ie a strong hardening of the spectrum. For example, X\thinspace 1624-490
shows a softening of the spectrun (Church and Balucinska-Church 1995),
and  X\thinspace 1755-338 shows energy independence (White et al. 1984;
Church and Balucinska-Church 1993).
We have previously proposed a physical model for the dipping sources which explains the spectral 
evolution in these two sources (Church and Balucinska-Church, 1995).
In this model, emission consists of two components:
blackbody emission from the neutron star plus extended Comptonised emission from the
accretion disk corona (ADC) represented as a power law at
energies well below the Comptonisation break. X\thinspace 1624-490 is also important
because dipping reaches a ``saturated'' lower level which is strong evidence
that two emission components are present (Church and Balucinska-Church
1995). The differences between the two sources
above are mainly due to differences in blackbody temperature. In X\thinspace 1624-490, 
$\rm {kT_{bb}}$ is relatively high, and as dipping
consists predominantly of absorption of the point-like blackbody emission, this
leaves the residual spectrum softer. In X\thinspace 1755-338, $\rm {kT_{bb}}$ is
smaller, with the peak emission at $\sim $ 3 keV, so that removal of the blackbody
leaves the spectrum neither harder nor softer in the band 1 - 10 keV.

\begin{figure*}[t]
\leavevmode\epsffile{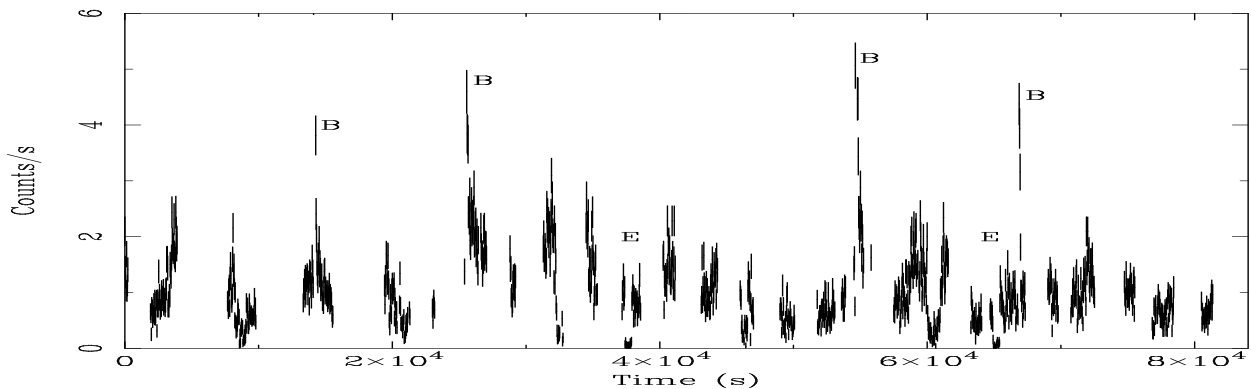}
\caption{ASCA GIS2 light curve for the complete 23~hr observation in the
energy band 0.7 - 10.0 keV with 32~s binning. To show the quiescent source more clearly,
the full heights of the bursts (B) have been suppressed. Eclipses are
labelled (E).\label{fig1}}
\end{figure*}

\medskip \noindent
There is also an important group of dipping sources including
XB\thinspace 0748-676, XB\thinspace 1916-053 and XB\thinspace 1254-690
which have very similar spectral evolution in dipping, in that there is clearly part of 
the spectrum which is not absorbed. These sources also, 
do not fit a simple pattern of photoelectric absorption in the bulge 
in the outer accretion disk. Spectral
evolution in dipping in these sources has been modelled using the
``absorbed plus unabsorbed'' approach (Parmar et al. 1986, Smale et al. 1988
and Courvoisier et al. 1986). In the case of XBT\thinspace 0748-676,
the non-dip spectrum was fitted with a cut-off power law model (Parmar et al.
1986). Dip data selected in intensity bands however, revealed a strong low energy 
excess, and the spectra could be fitted by dividing the non-dip model
into two parts with the same form, one strongly absorbed and the other not
absorbed. Good fits were obtained with the column density of the unabsorbed
part fixed at the non-dip value and but with strongly decreasing normalisation
as dipping progresses, whereas the absorbed component had increasing column
density but with a normalisation remaining at about the non-dip value. The origin
of the unabsorbed component has often in this group of sources been attributed
to electron scattering producing an energy-independent decrease of intensity,
although it is not clear that electron scattering should play an important role
in the bulge in the outer accretion disk in competition with photoelectric
absorption. 
In the case of XBT\thinspace 0748-676, Parmar et al. (1986) concluded that the
unabsorbed component was probably due to fast variations of column density
which could mimic a soft excess.

\medskip \noindent
In the case of XB\thinspace 1916-053, we have shown that the ASCA 
spectra may be fitted well by our two-component model including progressive
covering of the extended Comptonised emission (Church et al. 1997).
The blackbody emission from the neutron star was covered rapidly. Very good
fits were obtained in this way, with large increases in column density
$\rm {N_H}$ for the blackbody and smaller increases in $\rm {N_H}$ 
for the Comptonised emission, which are averages across the absorber.
The unabsorbed peak in the spectrum is the uncovered part of the power law
component, and the increase in the partial covering fraction from zero to
unity in deepest dipping is the behavior represented by the decreasing 
normalisation in ``absorbed + unabsorbed'' modelling. This is however radically 
different from that type of modelling. Firstly, we are explicitly 
using partial covering of extended emission in the model which has not previously
been done. Secondly, the emission model for the source has two terms, and is the
same model that we have shown gives good explanations of other dipping sources.
Dipping can be modelled well by photoelectric absorption alone, without the need for electron 
scattering, and calculations from the cross sections showed that scattering is
not expected to be important in the absorbing bulge in competition with
absorption below 10 keV (Church et al. 1997). The main aim of 
the present work was to test using the ASCA data whether this approach can also 
successfully describe spectral evolution in dipping in XBT\thinspace 0748-676.

\section{Observations}
We present results for the observation of XBT\thinspace 0748-676 made on 1993, May
7th with ASCA (Tanaka et al. 1994), which lasted 23 hours, during the performance 
verification phase. The GIS data were screened 
to remove regions of SAA passage, to restrict
elevation above the rim of the Earth to more than $\rm {5^{\circ}}$, particle rigidity 
to more than 6 GeV/c, the radiation belt parameter to less than 200 c/s and angular
deviation of the telescope pointing to less than $\rm {0.014^{\circ}}$. The calibration
source and outer ring were removed from the image, and rise-time rejection applied.
Source data were selected from a circle of radius 6\arcmin\ centered on the source. 
\placefigure{2}
Background data were also obtained from a circular region of radius
6\arcmin\ which
provided a sufficiently large background count to use for sensible subtraction from 
the spectra. SIS0 data were extracted similarly and `faint' data converted to
`bright' mode data. The data were screened, and cleaned to remove the effects of
hot pixels. The main spectral fitting results were however obtained by adding
GIS2 and GIS3 spectra to improve the quality of the spectra.

\begin{figure}[t]
\epsfxsize=90 mm
\leavevmode\epsffile{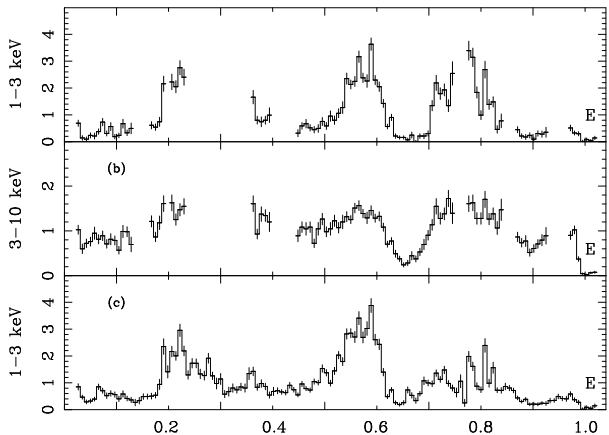}
\caption{ASCA GIS2 light curves with the 4 bursts removed folded on the orbital period
of 13,766.8~s: (a) in the band 1 - 3 keV, (b) in the band 3 - 10 keV and (c) from 
the total observation at 1- 3 keV. Eclipses are labelled (E).\label{fig2}}
\end{figure}

\section{Results}
The GIS2 light curve in the band 0.7 - 10.0 keV is shown in Fig. 1, in which
4 bursts and 2 eclipses can be seen. It is also evident that dipping reached 
a depth of 100\%, for example at $\sim $ 60,000~s. The source was faint
during the {\it ASCA} observation at $\sim $ 1 c/s; this would be about
1 c/s in the {\it Rosat PSPC}, compared with $\sim $ 30 c/s obtained in
the {\it PSPC} during the observation of 1991 Nov 11, for example
(Hertz et al. 1995). Thus the intensity
of the source had decreased by about a factor of 30. The orbital period was obtained 
by carefully measuring the times of eclipse ingress and egress from the GIS2
light curve with 4~s binning. Differences in time between eclipse centers gave
an orbital period P = $\rm {13,762\pm 6}$~s, consistent with the accepted
value of $\sim $ 13766.78~s. However, in producing our folded light curves, we
used the more accurate value. The bursts, but not the eclipses, were excluded from the 
data before folding. Of the 6 orbital cycles observed, only 2 were covered well. 
When all of the data is included, the folded light curve is distorted 
from the shape of individual dips; for example, the shape of the dips is affected
by the variability in the width of dipping. We therefore show in Fig. 2 folded 
light curves using orbits 3 and 5 which are not significantly distorted and 
which can be seen to be significantly different from the 
folded light curve including all data (lower panel). Note that in this figure, the dip between
phases 0.6 and 0.7 is only part of the main dipping activity which in this source generally
extends within an envelope between phases 0.5 and 1.1.
In the lowest energy band 1 - 3 keV, dipping extends
to almost all of the orbital cycle, suggesting the presence of absorber
spreading around the outer accretion disk. The dipping at phase
0.6 - 0.7 is 100\% deep and relatively steep-sided compared with dipping in the
band 3 - 10 keV which is only 80\% deep. Thus there is a marked increase in
hardness in this source, which contrasts with XB\thinspace 1916-053
during the {\it ASCA} observation, in which
the depth of dipping was $\sim $ 100\% at all energies in the band 1 - 10 keV
(Church et al. 1997). The eclipse in the folded light curve is clear in the center
panel in the band 3 -- 10 keV, but is less obvious in the band 1 -- 3 keV 
because dipping is in progress when the eclipse starts.

\medskip \noindent
GIS spectra were derived using the complete observation of 6 orbital cycles
as there was little evidence for dependence of the results obtained on 
whether all 6 orbits were used, or the first 3, or the last 3. Firstly,
GIS2 spectra were accumulated by removing bursts and eclipses and dividing 
the data into  intensity bins using the ranges: 2.4 - 3.2 c/s (non-dip), 
1.8 - 2.4 c/s, 1.2 - 1.8 c/s, 0.6 - 1.2 and 0.0 - 0.6 c/s. A consequence of
dipping extending to several parts of the orbital cycle was that little non-dip
data was available, and care was necessary to identify the correct intensity
band for non-dip to avoid including dip data. The folded light
curve
in the higher band 3 - 10 keV has a shape more typical of the dipping sources
and so was used to determine which data was non-dip. GIS2 and GIS3 data
could not be selected in the same intenity bands because of count rate
differences due to their different offsets from the bore-sight axis, and so
GIS3 data were selected via time filters corresponding to intensity
selections in GIS2. The GIS2 and GIS3 spectra 
were added, systematic errors of 2\% were applied conservatively to each channel, 
and the data regrouped to an appropriate minimum number of counts in each bin. 
Channels below 0.8 keV and above 10.0 keV were ignored. An average response
function was derived from the GIS2 and GIS3 responses appropriate to the source
position in each detector. Finally the GIS2 and GIS3 background 
spectra were added.  The non-dip data is of relatively poor
quality because of the predominance of dipping during the orbital cycle. In our
work on other dipping sources, it was possible to derive well-constrained
values of $\rm {kT_{bb}}$ and $\Gamma $ for our two-component model from the
non-dip data, and to fix these in fitting dip data. In this case, we have 
carried out simultaneous fitting of non-dip data and the 4 dip spectra.
The non-dip spectrum could be fitted by a simple absorbed power law adequately, 
although this model will not fit the dip spectra because of the absorbed and 
unabsorbed peaks in the spectrum. Simple thermal models are not capable of fitting 
the spectra. 

\medskip \noindent
Thus we next used the two-component
model which we have proposed for the dipping sources in general, consisting of
a blackbody component from the neutron star plus Comptonised emission
from the ADC with progressive covering of the 
ADC: $\rm {AB_1*BB\;+}$\hfill \break $\rm {AG*(AB_2*f\;+\; (1-f))*PL}$ 
where BB and PL are the blackbody and power law 
terms, AG is a Galactic absorption term, AB are variable absorption terms (including
the Galactic term for the blackbody), and {\it f} is the covering fraction.
The source emission parameters cannot change during dipping and so we require $\rm {kT_{bb}}$,
$\Gamma $ and the two normalisations to remain constant during dipping. 
Good fits to the 5 spectra simultaneously were obtained, with $\chi ^2$ for the simultaneous fitting of 312
for 517 degrees of freedom, suggesting that the systematic errors added
were pessimistic.

\medskip \noindent
We have also carried out spectral fitting of the SIS0 data in the energy band 0.5 - 10.0 keV
using the same spectral model. The model fitted the non-dip and dip data well at all 
energies above 1 keV, but at $\sim $ 0.7 keV, the soft excess reported by 
Thomas et al. (1997) was seen. First, we added an absorbed  Gaussian line to the
spectral model above, but found that this could only fit the spectra simultaneously
with a decreasing normalisation. This is exactly the behavior in the continuum emission
modelled in the absorbed + unabsorbed approach by a decreasing normalisation, but in our fitting
by the partial covering term. Consequently we next fitted a model in which a Gaussian
line was added to the power law term, thus applying the same partial covering fraction
to the line as to the continuum. This model was clearly able to fit the intensity
variation of the excess in dipping adequately.
We further found that no improvement in the fit was obtained by
allowing the line normalisation to vary in addition to the effect of the partial
covering term. This strongly suggests
that the excess originates in the same region as the Comptonised emission, ie in the 
ADC. The best fit energy from fitting was 0.65 keV,
which is the energy of O VIII L$\alpha$, although there are several lines
within $\sim $ 0.1 keV of this energy.

\begin{figure}[!ht]
\epsfxsize=90 mm  
\leavevmode\epsffile{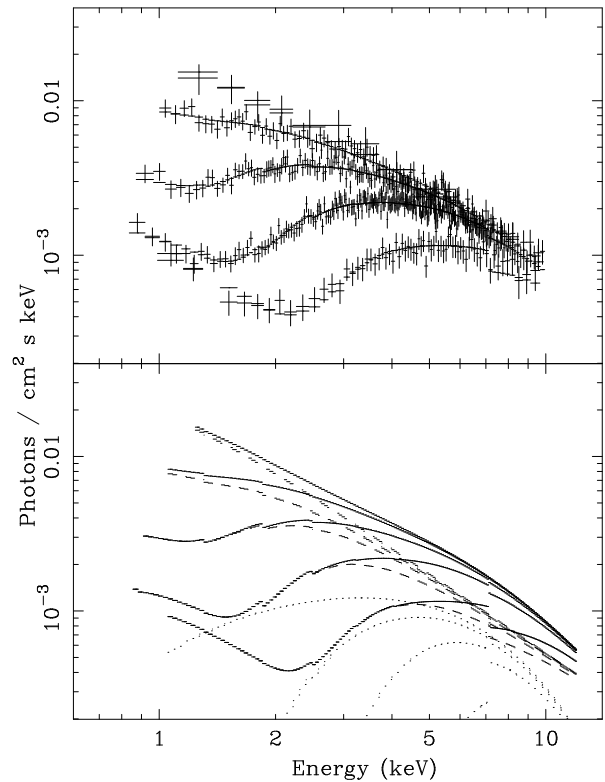}
\caption{Upper panel: fits to ASCA GIS2+GIS3 spectra for non-dip and several dip levels;
lower panel: the individual model components shown separately for clarity,
the blackbody is shown by dots, the power law
by dashes and the total model by a solid line.\label{fig3}}
\end{figure}

\medskip \noindent
Our best fit spectral fitting results from simultaneous fitting the 5 added GIS2 
and GIS3 spectra gave $\rm {kT_{bb}}$ = 1.99 $\pm $ 0.16 keV and power law photon
index $\Gamma $ = 1.70 $\pm $ 0.16. The value of the Galactic column density 
was $\rm {0.16 \pm 0.08 \cdot 10^{22}}$
H atom $\rm {cm^{-2}}$, in reasonable agreement with the Stark et al. value (1992)
of $\rm {0.1\cdot 10^{22}}$ H atom $\rm {cm^{-2}}$, and in good agreement with the
value of Parmar et al. (1986) of $\rm {0.18\cdot 10^{22}}$ H atom $\rm {cm^{-2}}$.
The best fit to non-dip and dip data is show in Fig. 3 and Table 1.

\placetable{1}
\begin{table}
\caption{Best fit spectral fitting results.
\label{1}}
\begin{tabular} {llrll}
$\rm {I^a}$ & $\rm {N_H(BB)^b}$ &$\rm {N_H(PL)^b}$ & f \\
\tableline
2.4 - 3.2  & $\sim $ 0.2    & 0.16$\pm $0.08  & 0.0             \\
1.8 - 2.4  & $\sim $ 0.2    & 1.6$\pm  $0.5   & 0.57$\pm $0.08  \\
1.2 - 1.8  & 4.8$\pm $1.5     & 2.1$\pm  $0.2    & 0.84$\pm $0.03   \\
0.6 - 1.2  & 17.7$\pm $2.7    & 4.7$\pm  $0.2  & 0.93$\pm $0.07   \\
0.0 - 0.6  & 79$\pm $16    & 12.3$\pm $0.8   & 0.95$\pm $0.07   \\
\tableline
\end{tabular}

$\rm {^a}$ intensity in c/s; $\rm {^b}$ in units of $\rm {10^{22}}$
H atom $\rm {cm^{-2}}$
\end{table}

\medskip \noindent
From Table 1 and Fig. 3, it can be seen that the spectral evolution in dipping was 
fitted by large increases in $\rm {N_H}$
for the blackbody, togther with smaller increases in $\rm {N_H}$ for the power law as this
was progressively covered by absorber. The blackbody is expected to measure 
$\rm {N_H}$ along a track across the absorber towards its center where the density is high,
whereas the smaller $\rm {N_H}$ of the power law suggests that $\rm {N_H}$ is
being integrated across the density gradient in the absorber.
In deepest dipping for which we have a spectrum, the blackbody is completely absorbed
and the partial covering fraction has risen to 95\%. 
However it is clear that the unabsorbed peak is the 5\% of incident
power law emission not covered by the absorber whereas the higher energy peak is the
residual of the power law covered by absorber after photoelectric absorption.
It is interesting that the blackbody in the first dip spectrum
persistently required an $\rm {N_H}$ of zero in addition to the Galactic contribution.
This supports the idea that $\sim $ 50\% of the extended emission region is covered by 
absorber before blackbody emission from the neutron star begins to be absorbed, as would be expected.
\placefigure{3}

\section{Discussion}
The two-component model which we have previously proposed for the dipping sources
has been shown to fit the source XBT\thinspace 0748-676. In addition, we have shown that
the spectral evolution in dipping can be explained in the same way as we did for 
XB\thinspace 1916-053, by allowing the extended Comptonised component to be progressively
covered by the absorber. In a separate analysis of the PV phase data for 
XBT\thinspace 0748-676, Thomas et al. (1997) use the absorbed + unabsorbed approach for
dip data, and find for non-dip emission that a single component generalised thermal model is 
`preferred from selected models'; however they do not report results using 
our two-component model, so their results should be seen in this light. 
Our spectral fitting
results give a total absorbed flux of $\rm {1.94\cdot 10^{-10}}$ erg $\rm {cm^{-2}}$ 
$\rm {s^{-1}}$ in the band 1 - 10 keV, of 
which the blackbody contribution is 29.9\%. The luminosity of the blackbody emission
calculated from the unabsorbed flux in the band 0.01 to 50 keV, and 
assuming a distance of 10 kpc (Gottwald et al. 1986) is 
$\rm {9.5\cdot 10^{35}}$ erg $\rm {s^{-1}}$, and from this,
we can calculate the height of the
emitting region assuming it is an equatorial strip around the star. This is found
to be very small: 0.09 km, indicative of a very thin accretion disk, consistent with 
the fading of the source by a factor of about 30 from the {\it Rosat}
observation of 1991 Nov 11.
In the case that the magnetic field strength of the neutron star were non-zero, 
the emitting area at the magnetic poles would still 
be very small: 5.8 $\rm {km^2}$.

The duration of eclipse or dip ingress and egress can in principle be used to derive the size of 
the extended source region, ie the ADC. However, in the present observation the source was faint, much
fainter than during the {\it Exosat} observations, and we have only two dips that are well
covered and 2 eclipses, compared with the 9 eclipses used by Parmar et al. (1986) to obtain
the duration of eclipse ingress. The count rate of the source in {\it RXTE}
was also much higher than in {\it ASCA} (Hertz et al. 1997).
In these circumstances we can only estimate the size of the ADC.
Using individual dips, we get an approximate duration of ingress $\rm {\Delta t}$ of about 110~s.
In the present case, the angular size of the absorber is known to be greater than that of the source
because dipping is 100\% deep
and so the duration of dip ingress is the time taken for the bulge in the outer disk to cross the
diameter of the ADC. We estimate the radius of the disk as $\rm {3.4\cdot 10^{10}}$~cm, and
the velocity of the bulge is $\rm {2\pi r_{disk}/P}$, where P is the orbital period.
From these we derive the radius of the ADC to be $\rm {8.5\cdot 10^{8}}$ cm. From the 2
eclipses, we find that the duration of ingress is between 16~s and 64~s. Combining these
with a binary separation of the stars of $\rm {1\cdot 10^{11}}$ cm, we get a radius of the ADC of
between $\rm{3.9\cdot 10^8}$ cm and $\rm {1.5\cdot 10^9}$ cm. 
This can be compared with the results from {\it Exosat} as follows.
Fig.~3 of Parmar et al. (1986) shows the profile of 9 eclipses added by folding on the orbital period,
having a steep transition to eclipse and also shoulders to the transition. The total duration
of ingress can be seen to be about 30~s, equivalent to a $\rm {r_{ADC}}$ of 
$\rm {7.2\cdot 10^8}$ cm. Our eclipse ingress times are consistent with
their value. Thus our dip and eclipse data suggest a value of $\rm {r_{ADC}}$ between
$\rm{3.9\cdot 10^8}$ cm and $\rm{1.5\cdot 10^9}$ cm, consistent with an extended region 
above the inner part of the accretion disk.

The spectrum of XBT\thinspace 0748-676 is clearly dominated by
non-thermal emission which is identified with Comptonised emission of the
ADC, and we can derive some information about the ADC from our results.
Firstly, the value of the Comptonisation y-parameter is found to be $\sim
$ 1.5 from our power law index of 1.7 (eg Rybicki and Lightman 1979).
There is no evidence (or expectation) for the Comptonisation break being
less than 10 keV; however we can assume that the break energy is similar
to that we have determined for the related source XB\thinspace 1916-053
using the very broad band of 0.1 - 300 keV of BeppoSAX (Church et al. 1998), 
ie $\sim $ 80 keV. 
From this value, the electron temperature $\rm {kT_e}$
is $\sim $ 30 keV, from which we derive values in XBT\thinspace 0748-676
of the average optical depth and column
density in the ADC of 2.5 and $\rm {4\cdot 10^{24}}$ $\rm {cm^{-2}}$.
We have also shown that the low energy feature can be fitted by a
Gaussian line at 0.65 keV in non-dip and dip spectra, and in addition have
shown that it originates in the ADC, since it can be fitted by the same
progressive factor as the extended continuum emission. Future
identification of this line, perhaps as O VIII L$\alpha$ will, of course,
provide direct information on conditions in the ADC.
High energy data is clearly required for
XBT\thinspace 0748-676, together with higher resolution data below 1 keV
to reveal the nature of the line emission.

The explanation of spectral evolution in XBT\thinspace 0748-676 
is radically different from previously modelling of the source using the absorbed + 
unabsorbed approach. Previously partial covering has not explicitly been used in 
modelling, and fitting our two-component model including partial covering of the
extended component shows that the unabsorbed emission can be identified with
uncovered non-dip emission. Gradual covering is represented in absorbed +
unsabsorbed modelling by the decreasing normalisation often attributed to electron
scattering. 
We do not need to invoke electron scattering in the absorber, so that 
dipping can be explained by photoelectric absorption alone. In the case of
XB\thinspace 1916-053 we presented calculations showing that electron scattering
is not expected to be important in the bulge in the outer disk at energies below 10 keV
(Church et al. 1997). In XBT\thinspace 0748-676, the density increase in dipping is less,
and so the effects of scattering will be even smaller.

\acknowledgments
MJC and MBC thank the Royal Society and the British Council for financial support.

\end{document}